\documentclass[reqno]{amsart}
\usepackage{amssymb}
\usepackage{verbatim}

\begin{document}

\title[Alternating group and multivariate exponential
functions]
{Alternating group and multivariate exponential functions}

\author{A.~Klimyk}
\address{Bogolyubov Institute for Theoretical Physics,
         Kiev 03680, Ukraine}
 \email{aklimyk@bitp.kiev.ua}

\author{J.~Patera}
\address{Centre de Recherches Math\'ematiques,
         Universit\'e de Montr\'eal,
         C.P.6128-Centre ville,
         Montr\'eal, H3C\,3J7, Qu\'ebec, Canada}
\email{patera@crm.umontreal.ca}

 \begin{abstract}
We define and study multivariate exponential functions, symmetric
with respect to the alternating group $A_n$, which is a subgroup
of the permutation (symmetric) group $S_n$. These functions are
connected with multivariate exponential functions, determined as
the determinants of matrices whose entries are exponential
functions of one variable. Our functions are eigenfunctions of the
Laplace operator. By means of alternating multivariate exponential
functions three types of Fourier transforms are constructed:
expansions into corresponding Fourier series, integral Fourier
transforms, and multivariate finite Fourier transforms.
Alternating multivariate exponential functions are used as a
kernel in all these Fourier transforms. Eigenfunctions of the
integral Fourier transforms are obtained.

 \end{abstract}

\maketitle

\noindent
2000 Math. Subject Classif.: Primary 42B05;
Secondary 20B30, 33E99, 42B10

\section{Introduction}
Mathematical and theoretical physics regularly deal with functions
on the Euclidean space $E_n$ that are symmetric or antisymmetric
with respect to the permutation (symmetric) group $S_n$. For
example, such functions describe collections of identical
particles. Symmetric and antisymmetric solutions appear in the
theory of integrable systems.

The symmetric group $S_n$ contains the alternating group $A_n$. It
is an invariant subgroup of $S_n$ of index 2 (that is, the group
$S_n/A_n$ has 2 elements). The alternating group $A_n$ consists of
transformations $w$ of the Euclidean space $E_n$ with $\det w=1$
and, therefore, is a subgroup of the rotation group $SO(n)$ (note
that $S_n$ does not belong to $SO(n)$). The group $A_n$ is simple.
For studying multivariate exponential functions, the group $A_n$
is more fundamental than the symmetric group, since multivariate
exponential functions symmetric or antisymmetric with respect to
$S_n$ can be constructed by means of the multivariate exponential
functions symmetric with respect to $A_n$.

The aim of this paper is to describe and study multivariate
exponential functions symmetrized by the alternating group and the
corresponding Fourier transforms. We call these functions {\it
alternating multivariate exponential functions} and denote them by
$E_\lambda(x)$, $\lambda=(\lambda_1,\lambda_2, \dots,\lambda_n)$,
$x=(x_1,x_2,\dots,x_n))\in E_n$. Such a function is a sum of terms
in the determinant of an $n\times n$ matrix (whose entries are
usual exponential functions of one variable) which enter in the
expression for the determinant with sign +. We call this sum a
semideterminant.

Alternating multivariate exponential functions are connected with
symmetric and antisymmetric multivariate exponential functions
studied in \cite{KP-07}. Symmetric and antisymmetric multivariate
exponential functions can be considered as a generalization of
cosine and sine functions of one variable, respectively, whereas
alternating multivariate exponential functions are a
generalization of the usual exponential function of one variable.
The connection of symmetric and antisymmetric multivariate
exponential functions with alternating multivariate exponential
functions is the same as that of the cosine and sine functions
with the exponential function of one variable (see section 2
below).

As in the case of the exponential functions of one variable, we may
consider three types of alternating multivariate exponential
functions:
 \medskip

(a) functions $E_m(x)$ with $m=(m_1,m_2,
\dots,m_n)$, $m_i\in {\mathbb{Z}}$, which determine Fourier
series expansions in alternating multivariate exponential
functions;
 \medskip

(b) functions $E_\lambda(x)$ with $\lambda=(\lambda_1,\lambda_2,
\dots,\lambda_n)$, $\lambda_i\in {\mathbb{R}}$, which determine
integral multivariate Fourier transforms;
 \medskip

(c) functions $E_\lambda(x)$, where $x=(x_1,x_2,
\dots,x_n)$ take a finite set of values; they determine multivariate
finite Fourier transforms.
 \medskip

 Functions (b) are symmetric with respect to elements
of the alternating group $A_n$.
Since exponential functions
$e^{2\pi{\rm i}mx}$, $m\in {\mathbb{Z}}$, of one variable
are invariant with respect to shifts $x\to x+k$, $k\in {\mathbb{Z}}$,
then symmetries of functions (a) are described by a wider group, which
is called the {\it affine alternating group} $A_n^{\rm aff}$. This group
is a product of the group $A_n$ and the group $T_n$, consisting of
shifts in the space $E_n$ by
vectors $r=(r_1,r_2,\dots,r_n)$, $r_j\in {\mathbb{Z}}$.
A fundamental domain $F(A_n^{\rm aff})$ of the group $A_n^{\rm aff}$
is a certain bounded subset of ${\mathbb{R}}^n$.

Functions on the fundamental domain $F(A_n^{\rm aff})$ can be
expanded into series in the functions (a). These expansions are an
analogue of the usual Fourier series for functions of one
variable. Functions (b) determine Fourier integral transforms on
the fundamental domain $F(A_n)$ of the alternating group $A_n$.
This domain consists of points $x\in E_n$ such that
$x_1,x_2>x_3>\cdots >x_n$, where $x_1,x_2>x_3$ means that
$x_1>x_3$ and $x_2>x_3$.

Functions (c) are used to determine finite (on a finite set)
Fourier transforms. These Fourier transforms are given on grids
consisting of points in the fundamental domain $F(A_n^{\rm aff})$.

Alternating multivariate exponential functions are closely related
to symmetric and antisymmetric exponential functions of
\cite{KP-07}. The symmetric and antisymmetric exponential
functions are connected with symmetric and antisymmetric orbit
functions defined in \cite{P04}, \cite{P-SIG-05} and studied in
detail in \cite{KP06} and \cite{KP07}. Discrete orbit function
transforms, corresponding to Coxeter--Dynkin diagrams of low
order, were studied in detail, and it was shown that they are very
useful for applications \cite{AP}--\cite{Appl-5}.

The discrete Fourier transforms, determined by (anti)symmetric
multivariate exponential functions, studied in \cite{KP-07}, and
by alternating multivariate exponential functions have a number of
practically useful properties. In particular, continuous extension
of the discrete transforms smoothly interpolate digital data in
any dimension. Examples show that relative to the amount of
available data, these transforms provide much smoother
interpolation than the conventional Fourier transforms.

Symmetric and antisymmetric multivariate exponential functions,
studied in \cite{KP-07}, satisfy certain boundary conditions
(antisymmetric exponential functions vanish on the boundary of the
corresponding fundamental domain and  the derivative of the
symmetric exponential functions with respect to the normal to the
boundary of the fundamental domain vanishes on the boundary). This
means that smooth functions, which are expanded in these
functions, have to satisfy these conditions, that is, not each
smooth function can be expanded in (anti)symmetric exponential
functions. Alternating multivariate exponential functions satisfy
no boundary conditions and any smooth function can be expanded in
these exponential functions.

Alternating multivariate exponential functions are also related to
the so-called $E$-orbit functions exposed in \cite{P04},
\cite{P-SIG-05}, and \cite{Ka-Pat}). The exposition of the theory
of orbit functions strongly depends on the theory of Weyl groups,
properties of root systems, etc. In this paper we avoid this
dependence. We use only the alternating group and its
affinization.

The best way to define alternating multivariate exponential
functions is to use a semideterminant of a finite matrix. The
semideterminant of a matrix is closely related to the determinant
and the antideterminant of the same matrix. It is well-known that
the determinant $\det (a_{ij})_{i,j=1}^n$ of an $n\times n$ matrix
$(a_{ij})_{i,j=1}^n$ is defined as
$$
\det (a_{ij})_{i,j=1}^n=\sum_{w\in S_n} (\det w)
a_{1,w(1)}a_{2,w(2)}\cdots a_{n,w(n)}
$$
\begin{equation}  \label{det}
= \sum_{w\in S_n} (\det w)
a_{w(1),1}a_{w(2),2}\cdots a_{w(n),n},
\end{equation}
where $S_n$ is the symmetric group of $n$ symbols $1,2,\dots,n$,
the set $(w(1),w(2),\dots$, $w(n))$ denotes the set
$w(1,2,\dots,n)$, and $\det w$ denotes the determinant of the
transform $w$, that is, $\det w=1$ if $w$ is an even permutation
and $\det w=-1$ otherwise. The antideterminant $\det^+$ of a
matrix $(a_{ij})_{i,j=1}^n$ is the sum of all terms, entering in
the expression for the corresponding determinant taken with  sign
+,
$$
{\det}^+ (a_{ij})_{i,j=1}^n=\sum_{w\in S_n} a_{1,w(1)}a_{2,w(2)}\cdots
a_{n,w(n)}= \sum_{w\in S_n}
a_{w(1),1}a_{w(2),2}\cdots a_{w(n),n}.
$$

For the semideterminant sdet of a matrix $(a_{ij})_{i,j=1}^n$ we
have
$$
{\rm sdet}\; (a_{ij})_{i,j=1}^n=
\frac12 \left( \det (a_{ij})_{i,j=1}^n+
{\det}^+ (a_{ij})_{i,j=1}^n \right).
$$
Clearly,
\begin{equation}  \label{dett}
{\rm sdet}\; (a_{ij})_{i,j=1}^n=
\sum_{w\in A_n} a_{1,w(1)}a_{2,w(2)}\cdots
a_{n,w(n)}= \sum_{w\in A_n}
a_{w(1),1}a_{w(2),2}\cdots a_{w(n),n}.
\end{equation}

In the text, we use formulas of the type
$$
N\ge a_1,a_2\ge a_3\ge a_4\ge \cdots \ge a_n.
$$
Here $N\ge a_1,a_2\ge a_3$ means that $N\ge a_1\ge a_3$ and $N\ge
a_2\ge a_3$. Formulas of the type $a_1,a_2>a_3>\cdots
>a_n$ hava a similar sense.

\section{Alternating multivariate exponential
functions}
An alternating multivariate exponential function $E_\lambda(x)$
of $x=(x_1,x_2$, $\dots,x_n)$ is defined as the function
\begin{align}  \label{sdetII}
E_\lambda(x)\equiv &
E_{(\lambda_{1},\lambda_2,\dots ,\lambda_{n})}(x)=
  {\rm sdet} \left( e^{2\pi{\rm
i}\lambda_ix_j}\right)_{i,j=1}^{n}
\notag\\
= & { {\rm sdet} \left(
 \begin{array}{cccc}
 e^{2\pi{\rm i}\lambda_1x_1}& e^{2\pi{\rm i}\lambda_1x_2}&\cdots & e^{2\pi{\rm
i}\lambda_1x_{n}}\\
  e^{2\pi{\rm i}\lambda_2x_1}& e^{2\pi{\rm i}\lambda_2x_2}&\cdots & e^{2\pi{\rm
i}\lambda_2x_{n}}\\
  \cdots & \cdots &
\cdots & \cdots \\
 e^{2\pi{\rm i}\lambda_{n}x_1}& e^{2\pi{\rm i}\lambda_{n}x_2}&\cdots & e^{2\pi{\rm
i}\lambda_{n}x_{n}} \end{array} \right) }
\notag\\
\equiv & \sum_{w\in A_n}
 e^{2\pi{\rm i}\lambda_1x_{w(1)}} e^{2\pi{\rm i}\lambda_2x_{w(2)}}\cdots  e^{2\pi{\rm
i}\lambda_nx_{w(n)}}
= \sum_{w\in A_n} e^{2\pi{\rm i}\langle \lambda, wx \rangle} ,
 \end{align}
where $\lambda=(\lambda_1,\lambda_2,\dots,\lambda_n)$ is a set of real
numbers, which determines the function $E_\lambda(x)$,
and $\langle \lambda,x  \rangle$ denotes the scalar product in the
$n$-dimensional Euclidean space $E_n$,
$\langle \lambda,x  \rangle=\sum_{i=1}^n\lambda_ix_i$.
When $\lambda_1,\lambda_2,\dots,\lambda_n$ are
integers, we denote this set of numbers as $m\equiv (m_1,m_2,\dots,m_n)$,
\begin{equation}   \label{detI}
E_{m_1,m_2,\dots,m_n}(x)= {\rm sdet} \left( e^{2\pi{\rm
i}m_ix_j}\right)_{i,j=1}^{n}.
 \end{equation}

{}From  expression \eqref{sdetII}, it follows that the alternating
exponential functions $E_\lambda(x)$ satisfy the relation
\begin{equation}  \label{x_1+x_2}
E_\lambda(x_1+a,x_2+a,\dots,x_n+a)=
e^{2\pi{\rm i}(\lambda_1+\lambda_2+\cdots +\lambda_n)a}
E_\lambda(x).
\end{equation}
It is therefore sufficient to consider the function $E_\lambda(x)$
on the hyperplane
\[
x_1+x_2+\cdots +x_n=b,
\]
where $b$ is a fixed number (we denote this hyperplane
by ${\mathcal{H}}_b$). A transition from one hyperplane
${\mathcal{H}}_b$ to another ${\mathcal{H}}_c$ is fulfilled by
multiplication by a usual exponential function
$e^{2\pi{\rm i}\vert\lambda \vert (c-b)}$, where
$\vert \lambda \vert=\lambda_1+\lambda_2+\cdots +\lambda_n$,
$$
E_\lambda(x)\vert_{x\in {\mathcal{H}}_b}=
e^{2\pi{\rm i}(\lambda_1+\lambda_2+\cdots +\lambda_n)(c-b)}
E_\lambda(x)\vert_{x\in {\mathcal{H}}_c}.
$$
It is useful to consider the functions
$E_\lambda(x)$ on the hyperplane
${\mathcal{H}}_0$.
For  $x\in {\mathcal{H}}_0$ we have the relation
\begin{equation}  \label{x_1-x_2}
E_{\lambda_1+\nu,\lambda_2+\nu,\dots,\lambda_n+\nu}(x)=
E_\lambda(x).
\end{equation}

We conclude from  \eqref{dett} that the expression of
semideterminant ${\rm sdet}$ does not change when applying a
permutation from $A_n$ to rows or to columns. This means that for
any permutation $w\in A_n$ we have
\begin{equation}  \label{permu}
E_{w\lambda}(x)=E_\lambda(x),\ \ \ \ E_{\lambda}(wx)=E_\lambda(x).
\end{equation}
Therefore, it is sufficient to consider only alternating
exponential functions $E_\lambda(x)$ with
$\lambda=(\lambda_1,\lambda_2,\dots,\lambda_n)$ such that
\[
\lambda_1,\lambda_2\ge \lambda_3\ge \cdots \ge \lambda_n.
\]
Such $\lambda$ are called {\it semidominant}. The set of all
semidominant $\lambda$ is denoted by $D^e_+$. Below, when
considering alternating exponential functions $E_\lambda(x)$, we
assume that $\lambda\in D^e_+$.

Alternating exponential functions are related to symmetric and
antisymmetric exponential functions $E^+_\lambda(x)$ and
$E^-_\lambda(x)$, which are studied in \cite{KP-07}. They are
determined by the formulas
$$
E^+_\lambda(x)=  {\rm det}^+ \left( e^{2\pi{\rm
i}\lambda_ix_j}\right)_{i,j=1}^{n},\ \ \ \
E^-_\lambda(x)=  {\rm det} \left( e^{2\pi{\rm
i}\lambda_ix_j}\right)_{i,j=1}^{n},
$$
where $\lambda$ and $x$ are such as in \eqref{sdetII}. This
relation will be considered in section 4. Here we consider the
case $n=2$.
 \medskip

{\bf The case $n=2$.} We examine the alternating multivariate
exponential functions $E_\lambda(x)$ for $n=2$ and
$\lambda_2=-\lambda_1\equiv \lambda$. Then
$$
E_{(\lambda,-\lambda)}(x_1,x_2)=e^{2\pi {\rm i}\lambda(x_1-x_2)},
\ \ \ \
E_{(-\lambda,\lambda)}(x_1,x_2)=e^{-2\pi {\rm i}\lambda(x_1-x_2)}.
$$
For antisymmetric and symmetric multivariate exponential functions
with these $n$ and $\lambda_2$ we have
$$
E^-_{(\lambda,-\lambda)}(x_1,x_2)
=e^{2\pi {\rm i}\lambda(x_1-x_2)}-e^{-2\pi {\rm i}\lambda(x_1-x_2)}
=2{\rm i}\sin 2\pi {\rm i}\lambda(x_1-x_2),
$$ $$
E^+_{(\lambda,-\lambda)}(x_1,x_2)=
e^{2\pi {\rm i}\lambda(x_1-x_2)}+e^{-2\pi {\rm i}\lambda(x_1-x_2)}
=2\cos 2\pi {\rm i}\lambda(x_1-x_2).
$$
Thus,
\begin{equation}\label{n=2}
E^-_{(\lambda,-\lambda)}(x_1,x_2)=
E_{(\lambda,-\lambda)}(x_1,x_2)-E_{(-\lambda,\lambda)}(x_1,x_2),
\end{equation}
\begin{equation}\label{n=2-}
E^+_{(\lambda,-\lambda)}(x_1,x_2)=
E_{(\lambda,-\lambda)}(x_1,x_2)+E_{(-\lambda,\lambda)}(x_1,x_2).
\end{equation}
Therefore, for $n=2$ and $\lambda_2=-\lambda_1$ the alternating
multivariate exponential function is the usual exponential
function of one variable, whereas antisymmetric and symmetric
multivariate exponential functions are sine and cosine functions
of one variable. As we shall see below, the relations \eqref{n=2}
and \eqref{n=2-} can be generalized for any $n\in \mathbb{Z}_+$.

\section{Affine alternating group and fundamental domains}
The functions $E_\lambda(x)$
are symmetric with respect to the alternating group $A_n$, that is,
$E_\lambda(wx)=E_\lambda(x)$, $w\in A_n$. The functions
$E_m(x)$ with integral $m=(m_1,m_2,\dots, m_n)$ admit
additional symmetries related to the periodicity of the exponential functions
$e^{2\pi{\rm i}r y}$, $r\in {\mathbb{Z}}$, $y\in {\mathbb{R}}$. These symmetries
are described by the discrete group of shifts in the space $E_n$ by vectors
\[
r_1{\bf e}_1+r_2{\bf e}_2+\cdots +r_n{\bf e}_n,\ \ \ r_i\in {\mathbb{Z}},
\]
where ${\bf e},{\bf e}_2,\dots, {\bf e}_n$ are the unit vectors
along the corresponding coordinate axes. We denote this group by
$T_n$.

Permutations of $A_n$ and shifts of $T_n$ generate a group which
is denoted as $A_n^{\rm aff}$ and referred to as the {\it affine
alternating group}. The group $A_n^{\rm aff}$ is a semidirect
product of its subgroups $A_n$ and $T_n$,
\[
A_n^{\rm aff}=A_n\, \circledS \;\, T_n,
\]
where $T_n$ is an invariant subgroup, that is, $wtw^{-1}\in T_n$ for $w\in A_n$
and $t\in T_n$.

An open connected simply connected set $F\subset {\mathbb{R}}^n$
is called a {\it fundamental domain} for the group $A_n^{\rm aff}$
(for the group $A_n$) if it does not contain equivalent points
(that is, points $x$ and $x'$ such that $x'=wx$, where $w$ belongs
to $A_n^{\rm aff}$ or $A_n$, respectively) and if its closure
contains at least one point from each $A_n^{\rm aff}$-orbit (from
each $A_n$-orbit). Recall that an $A_n^{\rm aff}$-orbit of a point
$x\in {\mathbb{R}}^n$ is the set of points $wx$, $w\in A_n^{\rm
aff}$.

It is evident that the set $D_{++}^e$ of all points $x=(x_1,x_2,\dots,x_n)$ such that
\[
x_1,x_2>x_2>\cdots >x_n,
\]
is a fundamental domain for the group $A_n$ (we denote it as
$F(A_n)$). The set of points $x=(x_1,x_2,\dots,x_n)\in D^e_{++}$
such that
\[
1>x_1,x_2>x_2>\cdots >x_n>0
\]
is a fundamental domain for the affine group $A^{\rm aff}_n$ (we
denote it as $F(A^{\rm aff}_n)$).

As previously seen, the functions $E_\lambda(x)$ are symmetric
with respect to the alternating group $A_n$. This means that it is
sufficient to consider the functions $E_\lambda(x)$ only on the
closure of the fundamental domain $F(A_n)$. Values of
$E_\lambda(x)$ on other points are obtained by using the symmetry.

The symmetry of functions $E_m(x)$, $m=(m_1,m_2,\dots,m_n)$,
$m_i\in \mathbb{Z}$, with respect to the affine alternating group
$A_n^{\rm aff}$,
\begin{equation}\label{aff-sym-aff}
E_m(wx+r)=E_m(x),\ \ \ \
w\in A_n,\ \ \ r\in T_n,
\end{equation}
means that we may consider
$E_m(x)$ only on the closure
of the fundamental domain $F(A^{\rm aff}_n)$, that is, on the set of points
$x$ such that
$1\ge x_1,x_2\ge x_2\ge \cdots \ge x_n\ge 0$.
Values of $E_m(x)$ on other points are obtained by
using the relation \eqref{aff-sym-aff}.

\section{Relation to symmetric and antisymmetric
exponential functions}
The alternating multivariate exponential functions
$E_\lambda(x)$ are related to
symmetric and antisymmetric multivariate exponential functions
$E^+_\lambda(x)$ and $E^-_\lambda(x)$ defined in \cite{KP-07} and determined
as
$$
E^-_\lambda(x)=\det \left( e^{2\pi {\rm i}\lambda_ix_j}\right)_{i,j=1}^n,
\ \ \ \
E^+_\lambda(x)={\det}^+ \left( e^{2\pi {\rm i}\lambda_ix_j}\right)_{i,j=1}^n,
$$
where ${\det}^+ \left( e^{2\pi {\rm i}\lambda_ix_j}\right)_{i,j=1}^n$ is
the antideterminant of the matrix
$\left( e^{2\pi {\rm i}\lambda_ix_j}\right)_{i,j=1}^n$.

It follows from the definitions of alternating and
symmetric and antisymmetric multivariate exponential functions
that for $\lambda$ such that
$\lambda_1>\lambda_2>\lambda_3> \cdots >\lambda_n$
we have
\begin{gather}\label{relat-1}
E^-_\lambda(x)=E_\lambda(x)-E_{r_{12}\lambda}(x),
\end{gather}
\begin{gather}\label{relat-2}
E^+_\lambda(x)=E_\lambda(x)+E_{r_{12}\lambda}(x),
\end{gather}
where $r_{12}$ means the permutation of $\lambda_1$ and $\lambda_2$.
It follows from here that
\begin{gather}\label{relat-3}   {\textstyle
E_\lambda(x)=\frac12 ( E^+_\lambda(x)+E^-_\lambda(x)), }
\end{gather}
\begin{gather}\label{relat-4}   {\textstyle
E_{r_{12}\lambda}(x)=\frac12 ( E^+_\lambda (x)-E^-_\lambda(x)). }
\end{gather}
It is directly derived from \eqref{relat-1}--\eqref{relat-4}
that
$$
(E^+_\lambda(x))^2-(E^-_\lambda(x))^2=4
E_{\lambda}(x)E_{r_{12}\lambda}(x),
$$  $$
(E^+_\lambda(x))^2+(E^-_\lambda(x))^2=
2(E_{\lambda}(x))^2+2(E_{r_{12}\lambda}(x))^2.
$$

If there are two coinciding numbers in the set
$\lambda_1,\lambda_2,\dots,\lambda_n$,  we get
\begin{gather}\label{relat-5}
E_\lambda(x)=E_\lambda^+(x).
\end{gather}

\section{Properties}
The symmetry of alternating multivariate exponential functions
$E_\lambda(x)$ with respect to the alternating group $A_n$ is a
main property of these functions. However, they possess many other
interesting properties.
 \medskip

{\bf Continuity.}
The functions $E_\lambda(x)$ are  finite sums of
multivariate exponential functions. Therefore, they are continuous
functions of $x_1,x_2,\dots,x_n$ and have
continuous derivatives of all orders in ${\mathbb{R}}^n$.

 \medskip

{\bf Complex conjugation.} Let there be two coinciding numbers in
the set $\lambda=(\lambda_1,\lambda_2,\dots ,\lambda_{n})$. Since
in this case $E_\lambda(x)=E_\lambda^+(x)$, then due to the
behavior of $E^+_\lambda(x)$ under complex conjugation (see
section 4 in \cite{KP-07}) we obtain
\begin{gather}\label{compl}
E_{(\lambda_{1},\lambda_2,\dots
,\lambda_{n})}(x)=\overline{E_{-(\lambda_{n},
\lambda_{n-1},\dots ,\lambda_{1})}(x)}.
\end{gather}
Let
$\lambda=(\lambda_1,\lambda_2,\dots ,\lambda_{n})$ be such that
$\lambda_1> \lambda_2> \cdots > \lambda_{n}$.
Then due to relations (17) and (18) in \cite{KP-07}
and formulas \eqref{relat-3} and \eqref{relat-4} we derive that
\begin{gather}\label{compl-0}
E_{(\lambda_{1},\lambda_2,\dots
,\lambda_{n})}(x)=\overline{E_{-(\lambda_{n},
\lambda_{n-1},\dots ,\lambda_{1})}(x)},
\end{gather}
\begin{gather}\label{compl-1}
E_{r_{12}(\lambda_{1},\lambda_2,\dots
,\lambda_{n})}(x)=\overline{E_{-r_{12}(\lambda_{n},
\lambda_{n-1},\dots ,\lambda_{1})}(x)}
\end{gather}
for $n=4k,4k+1$ and
\begin{gather}\label{compl-2}
E_{(\lambda_{1},\lambda_2,\dots
,\lambda_{n})}(x)=\overline{E_{-r_{12}(\lambda_{n},
\lambda_{n-1},\dots ,\lambda_{1})}(x)},
\end{gather}
\begin{gather}\label{compl-3}
E_{r_{12}(\lambda_{1},\lambda_2,\dots
,\lambda_{n})}(x)=\overline{E_{-(\lambda_{n},
\lambda_{n-1},\dots ,\lambda_{1})}(x)}
\end{gather}
for $n=4k-1,4k-2$, where $k$ is a positive integer.
 \medskip

{\bf Scaling symmetry.}
For $c\in {\mathbb{R}}$, let $c\lambda=(c\lambda_1,c\lambda_2,\dots,c\lambda_n)$.
Then
\[
E_{c\lambda}(x)=\sum_{w\in A_n} e^{2\pi{\rm i}\langle
cw\lambda ,x\rangle} = \sum_{w\in A_n}  e^{2\pi{\rm
i}\langle w\lambda ,cx\rangle} = E_{\lambda}(cx).
\]
The equality $E_{c\lambda}(x)=E_{\lambda}(cx)$
expresses the {\it scaling symmetry of exponential functions
$E_{\lambda}(x)$.}
 \medskip

{\bf Duality.}
Due to invariance of the scalar product $\langle \cdot, \cdot
\rangle$ with respect to the alternating group $A_n$, $\langle w\mu ,wy
\rangle= \langle \mu ,y \rangle$, we have
 \[
E_\lambda(x)=\sum_{w\in A_n} e^{2\pi {\rm i}\langle
\lambda,w^{-1}x \rangle} =\sum_{w\in A_n} e^{2\pi {\rm
i}\langle x, w\lambda \rangle}=  E_x(\lambda).
 \]
This relation expresses
the {\it duality} of alternating exponential functions.
 \medskip

{\bf Orthogonality on the fundamental domain $F(A_n^{\rm aff})$.}
Alternating exponential functions $E_m(x)$ with
$m=(m_1,m_2,\dots,m_n)\in D^e_{++}$, $m_j\in {\mathbb{Z}}$,
are orthogonal on $F(A_n^{\rm aff})$ with respect to the Euclidean
measure,
$$
|F(A_n^{\rm aff})|^{-1}\int_{\overline{F(A_n^{\rm aff})}}
E_m(x)\overline{ E_{m'}(x)}dx=
|A_n|\delta_{mm'} ,
$$
where $|A_n|$ means the number of elements in the set $A_n$,
$\overline{F(A_n^{\rm aff})}$ is the closure of $F(A_n^{\rm
aff})$, and $|F(A_n^{\rm aff})|$ is the area of the fundamental
domain $F(A_n^{\rm aff})$. This relation follows from the equality
 \begin{gather}\label{ortogo}
\int_{\sf T} E_m(x)
\overline{E_{m'}(x)}dx=
|A_n|\delta_{mm'}
  \end{gather}
(where ${\sf T}$ is the torus in $E_n$ consisting of points
$x=(x_1,x_2,\dots,x_n)$, $0\le x_i<1$), which is a consequence of
the orthogonality of the exponential functions $e^{2\pi{\rm
i}\langle \mu,x \rangle}$ (entering into the definition of
$E_m(x)$) for different sets $\mu$.

Assuming that an area of ${\sf T}$ is equal to 1, $|{\sf T}|=1$,
we have $|F(A_n^{\rm aff})|=|A_n|^{-1}$, and formula
\eqref{ortogo} takes the form
\begin{gather}\label{ortog-tor}
\int_{\overline{F(A_n^{\rm aff})}} E_m(x)
\overline{E_{m'}(x)}dx=
\delta_{mm'} .
 \end{gather}

There can be coinciding summands in expression \eqref{sdetII} for
alternating exponential functions. For this reason, for functions
$E_m(x)$ with $m=(m_1,m_2,\dots,m_n)\in D^e_+$, $m_j\in
{\mathbb{Z}}$, the relation \eqref{ortog-tor} is replaced by
\begin{gather}\label{ortog-torus}
\int_{\overline{F(A_n^{\rm aff})}} E_m(x)
\overline{E_{m'}(x)}dx=
|G_m| \delta_{mm'} .
 \end{gather}
where $|G_m|$ is the number of elements in the subgroup $G_m$ of
$A_n$ consisting of elements $w\in A_n$ such that $wm=m$.
 \medskip

{\bf Solutions of the Laplace equation.} The Laplace operator on
the Euclidean space $E_n$ in the Cartesian coordinates
$x=(x_1,x_2,\dots,x_n)$ takes the form
\[
\Delta=\frac{\partial^2}{\partial
x^2_1}+\frac{\partial^2}{\partial x^2_2}+\cdots
+\frac{\partial^2}{\partial x^2_n} .
\]
Taking any summand in the expression for the alternating
multivariate exponential function $E_\lambda(x)$, we get
\begin{gather*}
\Delta  e^{2\pi{\rm i}((w( \lambda))_1x_1+
 \cdots + (w(\lambda))_nx_{n})}
=-4\pi^2 \langle \lambda,\lambda \rangle\, e^{2\pi{\rm
i}((w( \lambda))_1x_1+
 \cdots + (w(\lambda))_nx_{n})},
\end{gather*}
where $\lambda=(\lambda_1,\lambda_2,\dots ,\lambda_n)$  determines
$E_\lambda(x)$.
Since this action of $\Delta$ does not depend on a summand
from the expression for the alternating
exponential function, we have
\begin{gather}\label{Lapl}
\Delta E_\lambda(x)= -4\pi^2\langle \lambda,\lambda \rangle
 E_\lambda(x),
 \end{gather}

The formula \eqref{Lapl} can be generalized in the following way.
Let $\sigma_k(y_1,y_2,\dots,y_n)$ be the $k$-th elementary
symmetric polynomial of degree $k$, that is,
\[
 \sigma_k(y_1,y_2,\dots,y_n)=\sum_{1\le k_1<k_2<\cdots <k_n\le n}
y_{k_1} y_{k_2}\cdots y_{k_n}.
\]
Then for $k=1,2,\dots,n$ we have
\begin{gather}\label{Lap-gene}
\sigma_k\left( \tfrac{\partial^2}{\partial
x^2_1},\tfrac{\partial^2}{\partial
x^2_2},\dots,\tfrac{\partial^2}{\partial x^2_n} \right)
E_\lambda(x)= (-4\pi^2)^k \sigma_k
(\lambda_1^2,\lambda_2^2,\dots,\lambda_n^2)
E_\lambda(x).
 \end{gather}
Note that $n$ differential equations \eqref{Lap-gene} are algebraically
independent.

Symmetric and antisymmetric multivariate exponential functions
also satisfy these equations. They satisfy the certain boundary
conditions (antisymmetric exponential functions vanish on the
boundary of the corresponding fundamental domain and  the
derivative of the symmetric exponential functions with respect to
the normal to the boundary of the fundamental domain vanishes on
the boundary). Alternating exponential functions do not satisfy
these conditions.

\section{Expansions in alternating
exponential functions on $F(A^{\rm aff}_n)$}

Alternating exponential functions determine symmetric (with
respect to $A_n$) multivariate Fourier transforms that generalize
the usual Fourier transform. There are three types of such
transforms:
 \medskip

(a) Fourier transforms related to the
functions $E_m(x)$ with $m=(m_1,m_2,\dots,m_n)$,
$m_j\in {\mathbb{Z}}$ (Fourier series);
 \medskip

(b) Fourier transforms related to $E_\lambda(x)$
with $\lambda\in D^e_+$;
 \medskip

(c) Multivariate finite Fourier transforms.
 \medskip

In this section, we consider expansions in alternating exponential
functions on the fundamental domain $F(A^{\rm aff}_n)$. These
expansions are constructed in the same way as in the case of
(anti)symmetric exponential functions in \cite{KP-07}.

Let $f(x)$ be a symmetric (with respect to the group $A_n^{\rm
aff}$) continuous function on the $n$-dimensional Euclidean space
$E_n$ which has continuous derivatives. We may consider this
function on the torus ${\sf T}$ which can be identified with a
closure of the union of the sets $w F(A^{\rm aff}_n)$, $w\in A_n$.
The function $f(x)$, as a function on ${\sf T}$, can be expanded
in exponential functions $e^{2\pi{\rm i}m_1 x_1} e^{2\pi{\rm i}m_2
x_2}\cdots e^{2\pi{\rm i}m_n x_n}$, $m_i\in {\mathbb{Z}}$. We have
 \begin{equation}\label{expan-anti}
f(x)=\sum_{m_i\in {\mathbb{Z}}} c_m
e^{2\pi{\rm i}m_1 x_1}
e^{2\pi{\rm i}m_2 x_2}\cdots e^{2\pi{\rm i}m_n x_n},
 \end{equation}
where $m=(m_1,m_2,\dots,m_n)$. It follows from the symmetry
$f(wx)=f(x)$, $w\in A_n$, that
\[
f(wx)=
\sum_{m_i\in {\mathbb{Z}}} c_m
e^{2\pi{\rm i}m_1 x_{w(1)}}
\cdots e^{2\pi{\rm i}m_n x_{w(n)}}
=\sum_{m_i\in {\mathbb{Z}}} c_{m}
e^{2\pi{\rm i}m_{w^{-1}(1)} x_{1}}
\cdots
e^{2\pi{\rm i}m_{w^{-1}(n)} x_n}
\] \[
=\sum_{m_i\in {\mathbb{Z}}} c_{wm}
e^{2\pi{\rm i}m_1 x_{1}}
\cdots e^{2\pi{\rm i}m_n x_{n}}
=f(x)=
\sum_{m_i\in {\mathbb{Z}}} c_m
e^{2\pi{\rm i}m_1 x_1}
\cdots e^{2\pi{\rm i}m_n x_n}.
\]
Therefore, the coefficients $c_m$ satisfy the conditions
$c_{wm}=c_m$, $w\in A_n$. Collecting exponential functions in
\eqref{expan-anti} at the same $c_{wm}$, $w\in A_n$, we obtain the
expansion
 \begin{equation}\label{expan-anti-2}
f(x)=\sum_{m\in P^e_+} c_m
E_m(x),
 \end{equation}
where $P^e_+=D^e_+\cap {\mathbb{Z}}^n$.
Thus, {\it any symmetric
(with respect to $A_n$) continuous function $f$ on ${\sf T}$
which has continuous derivatives (that is, any continuous function on $D^e_+$
with continuous derivatives) can be expanded in alternating
exponential functions} $E_m(x)$, $m\in P^e_+$.

By the orthogonality relation \eqref{ortog-torus}, the coefficients $c_m$ in the
expansion \eqref{expan-anti-2} are determined by the formula
 \begin{gather}\label{decom-ant-2}
c_m =|G_m|^{-1} \int_{\overline{F(A^{\rm aff}_n)}} f(x)
\overline{E_m(x)}dx ,
 \end{gather}
where, as before, $|G_m|$ is the number of elements in the
subgroup $G_m$ of $A_n$ consisting of $w\in A_n$ such that $wm=m$.
Moreover, the Plancherel formula holds:
 \begin{gather}\label{decom-ant-3}
\sum_{m\in P^e_+}  |c_m|^2=|G_m|^{-1}
\int_{\overline{F(A^{\rm aff}_n)}} |f(x)|^2dx.
 \end{gather}

Formula \eqref{decom-ant-2} is the symmetrized (with respect
to the group $A_n$) Fourier transform
of the function $f(x)$. Formula \eqref{expan-anti-2} gives an inverse
transform. Formulas \eqref{expan-anti-2} and \eqref{decom-ant-2} give the
{\it multivariate Fourier transforms} corresponding to the alternating
exponential functions $E_m(x)$, $m\in P^e_+$.

Let ${\mathcal L}^2(\overline{F(A^{\rm aff}_n)})$
denote the Hilbert space of functions on the domain
$\overline{F(A^{\rm aff}_n)}$ with the scalar product
\[
\langle f_1,f_2\rangle = \int_{\overline{F(A^{\rm aff}_n)}}
f_1(x)\overline{f_2(x)} dx .
\]
The formulas
\eqref{expan-anti-2}-\eqref{decom-ant-3} show that {\it the set of
alternating multivariate exponential
functions $E_m(x)$, $m\in P^e_+$, forms an
orthogonal basis of ${\mathcal L}^2(\overline{F(A^{\rm aff}_n)})$.}

\section{Fourier transforms on the fundamental domain
$F(A_n)$}
The expansions \eqref{expan-anti-2} of functions on the
fundamental domain $F(A^{\rm aff}_n)$ are expansions in the
exponential functions $E_m(x)$ with integral
$m=(m_1,m_2,\dots,m_n)$.  The exponential functions $E_\lambda(x)$
with $\lambda$ lying in the fundamental domain $F(A_n)$ (and not
obligatory integral) are not invariant with respect to the
corresponding affine alternating group $A_n^{\rm aff}$. They are
invariant  only with respect to the alternating group $A_n$. A
closure of the fundamental domain of $A_n$ coincides with the set
$D^e_+$ consisting of points $x$ such that $x_1,x_2\ge x_3\ge
\cdots \ge x_n$. The functions $E_\lambda(x)$, $\lambda\in D^e_+$,
determine Fourier transforms on $D^e_+$.

We begin with the usual Fourier transforms on ${\mathbb{R}}^n$:
 \begin{gather}\label{F-1}
\tilde f (\lambda)=\int_{{\mathbb{R}}^n} f(x) e^{2\pi {\rm
i}\langle \lambda,x \rangle} dx,
  \\
  \label{F-2}
 f (x)=\int_{{\mathbb{R}}^n} \tilde f(\lambda) e^{-2\pi {\rm i}\langle
\lambda,x \rangle} d\lambda.
  \end{gather}
Let the function $f(x)$ be invariant with respect to the
alternating group $A_n$, $f(wx)=f(x)$, $w\in A_n$. It is easy to
verify that $\tilde f (\lambda)$ is also invariant with respect to
$A_n$. By replacing $\lambda$ by $w\lambda$, $w\in A_n$, in
\eqref{F-1} and summing up both sides over $w\in A_n$, we obtain,
instead of \eqref{F-1}, the equality
 \begin{gather}\label{F-3}
\tilde f (\lambda)= \int_{D^e_+} f(x) E_\lambda(x) dx,\qquad
\lambda\in D^e_+,
  \end{gather}
where we have taken into account that $f(x)$ is invariant
with respect to $A_n$.

Similarly, starting from \eqref{F-2}, we obtain the inverse
formula:
 \begin{gather}\label{F-4}
 f (x)= \int_{D^e_+} \tilde f(\lambda)
 \overline{E_\lambda(x)} d\lambda .
  \end{gather}
For the transforms \eqref{F-3} and \eqref{F-4}, the Plancherel
formula
 \[
 \int_{D^e_+} |f(x)|^2 dx=
\int_{D^e_+} |\tilde f(\lambda) |^2  d\lambda
 \]
holds. Formulas \eqref{F-3} and \eqref{F-4} determine the
symmetric (with respect to $A_n$) multivariate Fourier transforms
on the domain $\overline{F(A_n)}$, which are referred to as the
{\it alternating Fourier transforms}.

Let ${\mathcal L}^2(D^e_+)$ denote the Hilbert space of functions on
the domain $D^e_+$ with the scalar product
\[
\langle f_1,f_2\rangle = \int_{D^e_+} f_1(x)\overline{f_2(x)} dx
\]
The formulas
\eqref{F-3} and \eqref{F-4} show that {\it the set of
alternating multivariate exponential
functions $E_\lambda(x)$, $\lambda\in D^e_+$, form a
full system of functions of ${\mathcal L}^2(D^e_+)$.}

\section{Eigenfunctions of the alternating Fourier transform}\label{section10}
Let $H_n(x)$, $n=0,1,2,\dots$, be the well-known Hermite polynomials of one
variable. They satisfy the relation
 \begin{equation}\label{Herm-1}
 \int_{-\infty}^\infty e^{2\pi {\rm i}px}e^{-\pi p^2}
 H_m(\sqrt{2\pi}p)dp={\rm
i}^{-m}e^{-\pi x^2}H_m(\sqrt{2\pi}x)
 \end{equation}
(see, for example, subsection 12.2.4 in \cite{KVII}).

We generate polynomials of many variables
\begin{equation}\label{Her-m}
H_{\bf m}({\bf  x})\equiv H_{m_1,m_2,\dots,m_n}(x_1,x_2,\dots
,x_n):=H_{m_1}(x_1)H_{m_2}(x_2)\cdots H_{m_n}(x_n).
 \end{equation}
The functions
\begin{equation}\label{Her-2}
 e^{-|{\bf  x}|^2/2} H_{\bf m}({\bf  x}),\ \ \ \
m_i=0,1,2,\dots,\ \ \ \ i=1,2,\dots,n,
 \end{equation}
where $|{\bf  x}|$ is a length of the vector ${\bf  x}$, form an
orthogonal basis of the Hilbert space ${\mathcal{L}}^2({\mathbb{R}}^n)$
with the scalar product
$\langle f_1,f_2\rangle:=\int_{{\mathbb{R}}^n}
f_1({\mathbf{x}})\overline{f_2({\mathbf{x}})}d{\mathbf{x}}$,
where $d{\mathbf{x}}=dx_1\,dx_2\cdots dx_n$.

We symmetrize the functions
\[
{\mathcal H}_{\bf m}({\bf x}):=e^{-\pi |{\bf x}|^2} H_{\bf m}(\sqrt{2\pi}
{\bf x})
\]
(obtained from \eqref{Her-2} by replacing ${\bf x}$ by
$\sqrt{2\pi}{\bf x}$) by means of alternating multivariate exponential functions:
\begin{equation}\label{Her-3}
\int_{{\mathbb{R}}^n} E_\lambda({\mathbf{x}})
 e^{-\pi|{\bf  x}|^2} H_{\bf m}(\sqrt{2\pi}{\bf  x})d{\bf  x}=
 {\rm i}^{-|{\bf m}|} e^{-\pi|\lambda|^2}
 {\bf H}_{\bf m}(\sqrt{2\pi}\lambda),
\end{equation}

It is easy to see that the polynomials ${\bf H}_{\bf m}$ are
indeed symmetric with respect to the group $A_n$,
 \[
{\bf H}_{\bf m}(w\lambda)={\bf H}_{\bf m}(\lambda),
\ \ \ \ w\in A_n.
\]
Therefore, we may consider ${\bf H}_{\bf m}(\lambda)$
for values of $\lambda=(\lambda_1,\lambda_2,\dots,\lambda_n)$
such that $\lambda_1,\lambda_2\ge \lambda_3\ge \cdots \ge \lambda_n$.
The polynomials ${\bf H}_{\bf m}$ are of the form
 \begin{equation}\label{sym-H}
{\bf H}_{\bf m}(\lambda)={\rm sdet}\
\left(H_{m_i}(\lambda_j)\right)_{i,j=1}^n,
 \end{equation}
that is, it is sufficient to consider the polynomials ${\bf
H}_{\bf m}(\lambda)$ for integer $n$-tuples ${\bf m}$, such that
$m_1,m_2\ge m_3\ge \cdots \ge m_n$.

Let us apply the alternating Fourier transform \eqref{F-3} (we denote it as
${\mathfrak{F}}$) to the functions \eqref{sym-H}. Taking into
account formula \eqref{Her-3} we derive
\begin{alignat*}{2}
{\mathfrak{F}}\left(  e^{-\pi|{\bf  x}|^2} {\bf H}_{\bf m}
(\sqrt{2\pi}{\bf x})\right)=& \; \frac1{|A_n|} \int_{{\mathbb{R}}^n}
E_\lambda({\mathbf{x}})
e^{-\pi|{\bf  x}|^2} {\bf H}_{\bf m}
(\sqrt{2\pi}{\bf x}) d{\bf x}
 \notag\\
=& \; {\rm i}^{-|{\bf m}|} e^{-\pi|\lambda|^2}
 {\bf H}_{\bf m}(\sqrt{2\pi}\lambda),
 \end{alignat*}
that is, the {\it functions $e^{-\pi|{\bf  x}|^2} {\bf H}_{\bf m}
(\sqrt{2\pi}{\bf x})$ are eigenfunctions of the alternating
Fou\-rier transform} ${\mathfrak{F}}$. Since these functions for
$m_i=0,1,2,\dots$, $i=1,2,\dots,n$, such that $m_1, m_2\ge m_3\ge
\cdots \ge m_n$, form an orthogonal basis of the Hilbert space
${\mathcal{L}}^2(D^e_+)$, they constitute a complete set of
eigenfunctions of this transform. Thus, this transform has only
four eigenvalues ${\rm i}, -{\rm i}, 1, -1$ in
${\mathcal{L}}^2(D^e_+)$. This means that we have
${\mathfrak{F}}^4=1$.

\section{Finite alternating Fourier transforms}
Along with the integral Fourier transform in one variable, there
exists a discrete Fourier transform in one variable. It is given
by the kernel
 \begin{equation}\label{f-F-1}
e_{mn}:=N^{-1/2}\exp (2\pi {\rm i}mn/N),\ \ \ \ m,n=1,2,\cdots,N,
 \end{equation}
where $N$ is a fix positive integer.
The matrix $(e_{mn})_{m,n=1}^N$ is unitary, that is,
 \begin{equation}\label{f-F-2}
\sum_k e_{mk}\overline{e_{nk}} =\delta_{mn},\ \ \ \ \sum_k
e_{km}\overline{e_{kn}} =\delta_{mn}.
 \end{equation}

Let $f(n)$ be a function of $n\in \{ 1,2\cdots ,N\}$. Then the function
 \begin{equation}\label{f-F-3}
\tilde f (m)=\sum_{n=1}^N f(n)e_{mn}\equiv N^{-1/2} \sum_{n=1}^N f(n) \exp
(2\pi{\rm i}mn/N)
 \end{equation}
is a finite Fourier transform of $f(n)$.
The inverse transform is given by
 \begin{equation}\label{f-F-4}
f(n)= N^{-1/2} \sum_{m=1}^N {\tilde f}(m) \exp (-2\pi{\rm i}mn/N).
 \end{equation}
The Plancherel formula
$\sum_{m=1}^N |\tilde f(m)|^2=\sum_{n=1}^N | f(n)|^2$
holds for transforms \eqref{f-F-3} and \eqref{f-F-4}. This means
that the finite Fourier transform conserves the norm introduced in
the space of functions on $\{ 1,2,\dots,N\}$.

In order to derive the finite alternating multivariate Fourier
transform, we use the discrete exponential function \eqref{f-F-1}
in the form
\begin{equation}\label{mult-F-1-1-1}
 \textstyle{
e_m(s):=N^{-1/2}\exp (2\pi{\rm i}ms),\ \ \ s\in F_N\equiv\{ \frac1N,
\frac2N,\dots,\frac{N-1}N ,1\},\ \ \ m\in {\mathbb Z}^{\ge 0},
}
\end{equation}
and generate a multivariate discrete exponential function by
taking a product of $n$ copies of these functions,
\begin{alignat}{2}\label{mult-F-1}
e_{\bf m}({\bf s}):=&\; e_{m_1}(s_1) e_{m_2}(s_2)\cdots
e_{m_n}(s_n)\notag\\
=&\; N^{-n/2}\exp (2\pi{\rm i}m_1s_1)\exp (2\pi{\rm i}m_2s_2)
\cdots \exp (2\pi{\rm i}m_ns_n)
 \end{alignat}
where ${\bf s}=(s_1,s_2,\dots,s_n)\in F^n_N$ and
${\bf m}=(m_1,m_2,\dots,m_n)\in ({\mathbb Z}^{\ge 0})^n$.
We introduce a scalar product in the space of linear combinations of
the functions \eqref{mult-F-1} by the formula
\begin{equation}\label{mult-F-3a}
\langle e_{\bf m}({\bf s}),e_{{\bf m}'}({\bf s})   \rangle
\equiv \prod_{i=1}^n \langle e_{m_i}(s_i), e_{m'_i}(s_i)\rangle
:=\prod_{i=1}^n \sum_{s_i\in F_N}e_{m_i}(s_i)\overline{e_{m'_i}(s_i)}
=\delta_{{\bf m}{\bf m}'}
\end{equation}
where $m_i,m_i'\in \{ 1,2,\dots, N\}$.
Here we used the relations \eqref{f-F-2}.

We now take  multivariate functions \eqref{mult-F-1} for integers
$m_i$ such that
\[
N\ge m_1,m_2\ge m_3\ge \cdots\ge m_n\ge 1
\]
and symmetrize them with respect to the alternating group $A_n$.
We obtain a finite version of the alternating exponential
functions \eqref{sdetII},
\begin{equation}\label{mult-F-10}
\tilde E_{\bf m}({\bf s}):=
|A_n|^{-1/2}  {\rm sdet}^+ (e_{m_i}(s_j))_{i,j=1}^n=
|A_n|^{-1/2}N^{-n/2} E_{\bf m}({\bf s}),
 \end{equation}
where the discrete functions $e_m(s)$ are given by
\eqref{mult-F-1-1-1}.

The $n$-tuples ${\bf s}$ in \eqref{mult-F-10} run over $F_N^n\equiv
F_N\times \cdots\times F_N$ ($n$ times).
We denote by $\breve F_N^n$ the subset of $F_N^n$ consisting of ${\bf s}=(s_1,s_2,\dots,s_n)\in
F_N^n$ such that
\[
 s_1,s_2\ge s_2\ge \cdots \ge s_n.
\]
The set $\breve F_N^n$ is a finite subset of the closure of the fundamental
domain $F(A^{\rm aff}_n)$.

Acting by permutations $w\in A_n$ upon $\breve F_N^n$ we obtain
the whole set $F_N^n$, where each point, having some coinciding
coordinates $m_i$, is repeated several times. Namely, a point
${\bf s}$ is contained $|G_{\bf s}|$ times in $\{ w\breve F_N^n;
w\in A_n\}$, where $G_{\bf s}$ is the subgroup of $A_n$ consisting
of elements $w\in A_n$ such that $w {\bf s}={\bf s}$.

By $\breve D_N^n$ we denote the set of integer $n$-tuples ${\bf m}=(m_1,m_2,\dots,m_n)$
such that
\[
 N\ge m_1,m_2\ge m_3\ge \cdots\ge m_n\ge 1.
\]

\noindent
{\bf Proposition.} {\it For ${\bf m},{\bf m}'\in \breve D_N^n$
the discrete functions \eqref{mult-F-10}
satisfy the orthogonality relation}
\begin{equation}\label{mult-F-11}
\langle \tilde E_{\bf m}({\bf s}),
\tilde E_{{\bf m}'}({\bf s})\rangle =
|A_n|\sum_{{\bf s}\in \breve F_M^n} |G_{\bf s}|^{-1}
\tilde E_{\bf m}({\bf s})
\overline{ \tilde E_{{\bf m}'}({\bf s})}
=|G_{\bf m}| \delta_{{\bf m}{\bf m}'}.
 \end{equation}

{\bf Proof.} Due to the definition of the scalar product we get
\begin{alignat*}{2}
\langle \tilde E_{\bf m}({\bf s}),
\tilde E_{{\bf m}'}({\bf s})\rangle =&\;
\sum_{{\bf s}\in F_N^n} \tilde E_{\bf m}({\bf s})
\overline{\tilde E_{{\bf m}'}({\bf s})} \\
 =&\; |A_n|^{-1}|G_{\bf m}| \sum_{w\in A_n} \prod_{i=1}^n
\sum_{s_i\in F_N} e_{m_{w(i)}}(s_i)\overline{e_{m'_{w(i)}}(s_i)}
= |G_{\bf m}| \delta_{{\bf m}{\bf m}'},
\end{alignat*}
where we have taken into account that some
summands in the expression for sdet in \eqref{mult-F-10} can coincide.

Since functions $\tilde E_{\bf m}({\bf s})$ are
symmetric with respect to $A_n$, then
\[
 \sum_{{\bf s}\in F_N^n} \tilde E_{\bf m}({\bf s})
\overline{\tilde E_{{\bf m}'}({\bf s})} =|A_n|
\sum_{{\bf s}\in \breve F_N^n} |G_{\bf s}|^{-1}
\tilde E_{\bf m}({\bf s})
\overline{\tilde E_{{\bf m}'}({\bf s})} ,
\]
where we have taken into account that under the action by $A_n$ upon
$\breve F_N^n$ a point ${\bf s}$ appears $|G_{\bf s}|$ times
in $F^n_N$.
This proves the proposition.
 \medskip

Let $f$ be a function on $\breve F^n_N$ (or a symmetric function on $F_N^n$).
Then it can be expanded in functions \eqref{mult-F-10} as
\begin{equation}\label{mult-F-12}
f({\bf s})=\sum_{{\bf m}\in \breve D_N^n}a_{\bf m}
\tilde E_{\bf m}({\bf s}).
 \end{equation}
The coefficients  $a_{\bf m}$ are determined by the formula
\begin{equation}\label{mult-F-13}
a_{\bf m}=|A_n| |G_{\bf m}|^{-1}\sum_{{\bf s}\in \breve F_N^n}
|G_{\bf s}|^{-1}f({\bf s}) \overline{\tilde E_{\bf m}({\bf s})}.
 \end{equation}

The expansions \eqref{mult-F-12} and \eqref{mult-F-13} follow from
the facts that numbers of elements in $\breve D_N^n$ and in
$\breve F_N^n$ are the same and from the orthogonality relation
\eqref{mult-F-11} (see also \cite{MP06}). We refer to expansions
\eqref{mult-F-12} and \eqref{mult-F-13} as the {\it alternating
multivariate finite Fourier transforms}.

\subsection*{Acknowledgements}
The research of the first author was partially supported by Grant
14.01/016 of the State Foundation of Fundamental Research of
Ukraine. We acknowledge also partial support for this work by the
National Science and Engineering Research Council of Canada,
MITACS, the MIND Research Institute.

\end{document}